\documentclass[12pt]{article}

\begin{document}

\input{epsf}

%%%%%%%%%%%%%%%%%%%%%%%%%%
% Vatche Sahakian's macros

\newcommand{\bb}{\begin{equation}}
\newcommand{\ee}{\end{equation}}
\newcommand{\bbb}{\begin{eqnarray}}
\newcommand{\eee}{\end{eqnarray}}
\newcommand{\vc}[1]{\mbox{$\vec{{\bf #1}}$}}
\newcommand{\mc}[1]{\mathcal{#1}}
\newcommand{\del}{\partial}
\newcommand{\lk}{\left}
\newcommand{\ave}[1]{\mbox{$\langle{#1}\rangle$}}
\newcommand{\re}{\right}
\newcommand{\pd}[1]{\frac{\del}{\del #1}}
\newcommand{\pdd}[2]{\frac{\del^2}{\del #1 \del #2}}
\newcommand{\Dd}[1]{\frac{d}{d #1}}
\newcommand{\sech}{\mbox{sech}}
\newcommand{\pref}[1]{(\ref{#1})}

\newcommand
{\sect}[1]{\vspace{20pt}{\LARGE}\noindent
{\bf #1:}}
\newcommand
{\subsect}[1]{\vspace{20pt}\hspace*{10pt}{\Large{$\bullet$}}\mbox{ }
{\bf #1}}
\newcommand
{\subsubsect}[1]{\hspace*{20pt}{\large{$\bullet$}}\mbox{ }
{\bf #1}}

\def\ie{{\it i.e.}}
\def\eg{{\it e.g.}}
\def\cf{{\it c.f.}}
\def\etal{{\it et.al.}}
\def\etc{{\it etc.}}

\def\AA{{\cal A}}
\def\BB{{\cal B}}
\def\CC{{\cal C}}
\def\DD{{\cal D}}
\def\EE{{\cal E}}
\def\FF{{\cal F}}
\def\GG{{\cal G}}
\def\HH{{\cal H}}
\def\II{{\cal I}}
\def\JJ{{\cal J}}
\def\KK{{\cal K}}
\def\LL{{\cal L}}
\def\MM{{\cal M}}
\def\NN{{\cal N}}
\def\OO{{\cal O}}
\def\PP{{\cal P}}
\def\QQ{{\cal Q}}
\def\RR{{\cal R}}
\def\SS{{\cal S}}
\def\TT{{\cal T}}
\def\UU{{\cal U}}
\def\VV{{\cal V}}
\def\WW{{\cal W}}
\def\XX{{\cal X}}
\def\YY{{\cal Y}}
\def\ZZ{{\cal Z}}

\def\sinh{{\rm sinh}}
\def\cosh{{\rm cosh}}
\def\tanh{{\rm tanh}}
\def\sgn{{\rm sgn}}
\def\det{{\rm det}}
\def\exp{{\rm exp}}
\def\sh{{\rm sh}}
\def\ch{{\rm ch}}

\def\ell{{\it l}}
\def\str{{\it str}}
\def\lp{\ell_{{\rm pl}}}
\def\blp{\overline{\ell}_{{\rm pl}}}
\def\ls{\ell_{{\str}}}
\def\bls{{\bar\ell}_{{\str}}}
\def\bM{{\overline{\rm M}}}
\def\eff{{\rm eff}}
\def\gs{g_\str}
\def\gym{g_{\sst\rm YM}}
\def\geff{g_{\eff}}
\def\r11{R_{11}}
\def\kel{{2\kappa_{11}^2}}
\def\kten{{2\kappa_{10}^2}}
\def\lpten{{\lp^{(10)}}}
\def\alp{\alpha'}
\def\aleff{{\alp_{\eff}}}
\def\sqaleff{{\alpha^{\prime\,1/2}_\eff}}
\def\tgs{{\tilde{g}_s}}
\def\talp{{{\tilde{\alpha}}'}}
\def\tlp{{\tilde{\ell}_{{\rm pl}}}}
\def\tr11{{\tilde{R}_{11}}}
\def\wtilde{\widetilde}
\def\what{\widehat}
\def\hlp{{\hat{\ell}_{{\rm pl}}}}
\def\hr11{{\hat{R}_{11}}}
\def\hf{{\textstyle\frac12}}
\def\coeff#1#2{{\textstyle{#1\over#2}}}
\def\CY{Calabi-Yau}
\def\lessapprox{\;{\buildrel{<}\over{\scriptstyle\sim}}\;}
\def\greaterapprox{\;{\buildrel{>}\over{\scriptstyle\sim}}\;}
\def\inbar{\,\vrule height1.5ex width.4pt depth0pt}
\def\IC{\relax\hbox{$\inbar\kern-.3em{\rm C}$}}
\def\IR{\relax{\rm I\kern-.18em R}}
\def\IP{\relax{\rm I\kern-.18em P}}
\def\Z{{\bf Z}}
\def\R{{\bf R}}
\def\One{{1\hskip -3pt {\rm l}}}
\def\sst{\scriptscriptstyle}
\def\osc{{\rm\sst osc}}
\def\lam{\lambda}
\def\lc{{\sst LC}}
\def\pr{{\sst \rm pr}}
\def\cl{{\sst \rm cl}}
\def\D{{\sst D}}
\def\bh{{\sst BH}}
\def\vev#1{\langle#1\rangle}

\def\Rtil{{\tilde R}}
\def\Vtil{{\tilde V}}

%%%%%%%%%%%%%%%%%%%%%%%%%%%%%%%%%%%%%%%%%%%%%%%%%%%%%%%%%%%%%%%%%%%

\begin{titlepage}
\rightline{EFI-99-28}

\rightline{hep-th/9906137}

\vskip 1cm
\begin{center}
\Large{{\bf 
A note on the thermodynamics\\ 
of `little string' theory}}
\end{center}

\vskip 1cm
\begin{center}
Emil Martinec\footnote{\texttt{ejm@theory.uchicago.edu}} ~~{\it and}~~ 
Vatche Sahakian\footnote{\texttt{isaak@theory.uchicago.edu}}
\end{center}
\vskip 12pt
\centerline{\sl Enrico Fermi Inst. and Dept. of Physics}
\centerline{\sl University of Chicago}
\centerline{\sl 5640 S. Ellis Ave., Chicago, IL 60637, USA}

\vskip 2cm

\begin{abstract}

We study the thermodynamics of the D1-D5 system on a five-torus, 
focussing on the roles of different scales. 
One can take a decoupling limit such that the tension
of the `little string' inside the fivebrane remains finite
and the physics is 5+1 dimensional.
The dual black geometry exhibits a boosted Hagedorn phase,
as well as a phase describing a boosted fivebrane gas.
The dependence on the boost yields information about the nature of the
fivebrane modes and their interactions.
In particular, the form of the equations of state suggests a description 
in terms of $k=Q_1Q_5$ degrees of freedom, 
which may lead to an explanation of the $Q_5^{\,3}$ growth 
in the fivebrane density of states below the Hagedorn transition.

\end{abstract}

\end{titlepage}
\newpage
\setcounter{page}{1}
\paragraph{Introduction:}
Maldacena's conjecture~\cite{MALDA1} 
has led to a systematic exploration of
strongly coupled dynamics of non-gravitational theories. Zero
temperature physics was studied in~\cite{GUBSER,WITHOLO}; some
thermodynamic aspects were considered in~\cite{MALDA2,WITPHASE};
the phase diagrams of super Yang-Mills (SYM)
and little string theories were charted in~\cite{RABIN,MSSYM123,MSFIVE}. 
The reader is referred to~\cite{ADSLECT} 
for a more complete list of 
references.  In this note, 
we extend our previous analysis \cite{MSFIVE} of
the D1-D5 system to 
demonstrate interesting features that are scaled away 
in the standard Maldacena limit of this system%
\footnote{This line of thought was instigated 
by questions and comments from O. Aharony in regard to our previous
paper~\cite{MSFIVE}.}.

The black geometry of a configuration of $Q_1$ D1 branes and $Q_5$ D5 branes
is dual to a non-gravitational theory with sixteen 
supercharges, a theory of little strings propagating
in six dimensions.  The theory is non-local on a scale
set by the little string tension.  We parameterize the
phase structure of this theory as in~\cite{MSFIVE} using 
some of the D1-D5 moduli: The six dimensional string coupling
$g_6$; the cycle size $R$ on which the $Q_1$ D1 branes are wrapped; the
string tension $\alp$ of the IIB theory; and the volume
of the additional (square) four-torus on which the D5 branes 
are wrapped, denoted by $V_4\equiv v \alp^2$.
The little string tension is then given by the 
't Hooft coupling of the D5 branes
\bb\label{efftension}
T_{ls}=\frac{1}{2\pi \aleff}=\frac{1}{2\pi Q_5 g_6 v^{1/2} \alp}
=\frac{(2\pi)^2}{Q_5 (\gym^{\sst (D5)})^2}\ ,
\ee
which is the energy cost per unit length of instanton strings
in the 5+1 gauge theory.
It was proposed in~\cite{SEIBLITTLE,ABKSS} that the limit of
decoupling from the gravitational bulk corresponds to an energy regime
set by this tension 
\bbb\label{seiblim}
\alp & \rightarrow & 0\ ,\mbox{with }g_6\ ,\ R\ ,\mbox{ and }v^{1/2}\alp
\mbox{ (hence }\aleff)\mbox{ held fixed.}
\eee
Thermodynamic phase diagrams of the theory
in this energy regime and with $Q_1=0$ were
studied in~\cite{MSFIVE} in the context of D4 and D5 branes. 
An alternative limit, proposed in~\cite{MALDA1},
was used in~\cite{MSFIVE} for the D1-D5 system
\bb\label{maldalimit}
\alp\rightarrow 0\ ,\mbox{with }g_6\ ,\ R\ ,\mbox{ and }v
\mbox{ held fixed.}
\ee
In this limit, the little string tension is scaled away, 
$\aleff\rightarrow 0$, and the system exhibits
1+1d conformal symmetry.
Yet a third energy regime can be identified, corresponding
to that of Matrix theory~\cite{MAT1} in the presence of a 
longitudinal fivebrane
\bb\label{matrixlimit}
\alp\rightarrow 0\ ,\mbox{with }g_6\ ,\ R\ ,\mbox{ and }v^{1/2}/\alp
\mbox{ held fixed,}
\ee
which implies the `DLCQ' limit
\bb
\lp\rightarrow 0\ ,\mbox{with }\frac{\lp^2}{\r11}\ ,\ \frac{\lp}{\RR}\ ;
\mbox{ and }\frac{\lp^4}{\VV_4}
\mbox{ held fixed,}
\ee
where $\lp$, $\VV_4$, $\RR$ and $\r11$ are moduli in a description
dual to the D1-D5 frame, via T-duality on the circle $R$
and lifting to M-theory.
This limit also sends the little string tension to infinity.

We encounter these three apparently different
energy regimes because this system has two
charges, and consequently two energy scales 
which can be taken as the appropriate powers 
of the SYM couplings
$\gym^{\sst(D1)}$ and $\gym^{\sst(D5)}$ of the two sets of branes.
The three regimes differ only in how
the volume of the four-torus $V_4$ is arranged with respect to the
string scale: 
in~\pref{seiblim}, $V_4$ is held fixed (\ie\ $V_4\gg \alp^2$); 
$V_4\sim \alp^2$ in the second case~\pref{maldalimit}; 
and $V_4\sim 1/\alp^2$ in the third
scenario~\pref{matrixlimit} 
(\ie\ $V_4\ll \alp^2$).  We discuss in detail
the energy regime~\pref{seiblim}, 
where the little string tension is held fixed; 
we will then comment only briefly on the roles of
the other two regimes, as no new physics arises in these cases
(for a discussion of the regime~\pref{maldalimit}, 
the reader may also consult~\cite{MSFIVE}).
It is useful to define the effective dual circle radius
\bb\label{dualr}
  \Rtil=\frac{\aleff}{R}
\ee
and the effective transverse box size 
(transverse volume per instanton string)
\bb\label{veff}
  \Vtil=\left(\frac{Q_5V_4}{Q_1}\right)^{1/4}\ ,
\ee
since these are the dimensionful quantities that, together with $\aleff$,
parametrize all the strong coupling equations of state.
We will treat the little string tension
$\aleff$ as the basic scale in the theory, referring
all other dimensionful quantities 
(such as the cycle sizes $\Rtil$ and $\Vtil$) to this scale.
Phase diagrams are plotted for fixed $\Rtil/\sqaleff$
as a function of $S$ and $\Vtil/\sqaleff$.
The phase diagram differs qualitatively for 
$\Rtil\gg\sqaleff$ and $\Rtil\ll\sqaleff$.
We use the same notation as in~\cite{MSFIVE}, in particular
$k\equiv Q_1 Q_5$ and $q=Q_1/Q_5>1$.

\begin{figure}
\epsfxsize=13.5cm \centerline{\leavevmode \epsfbox{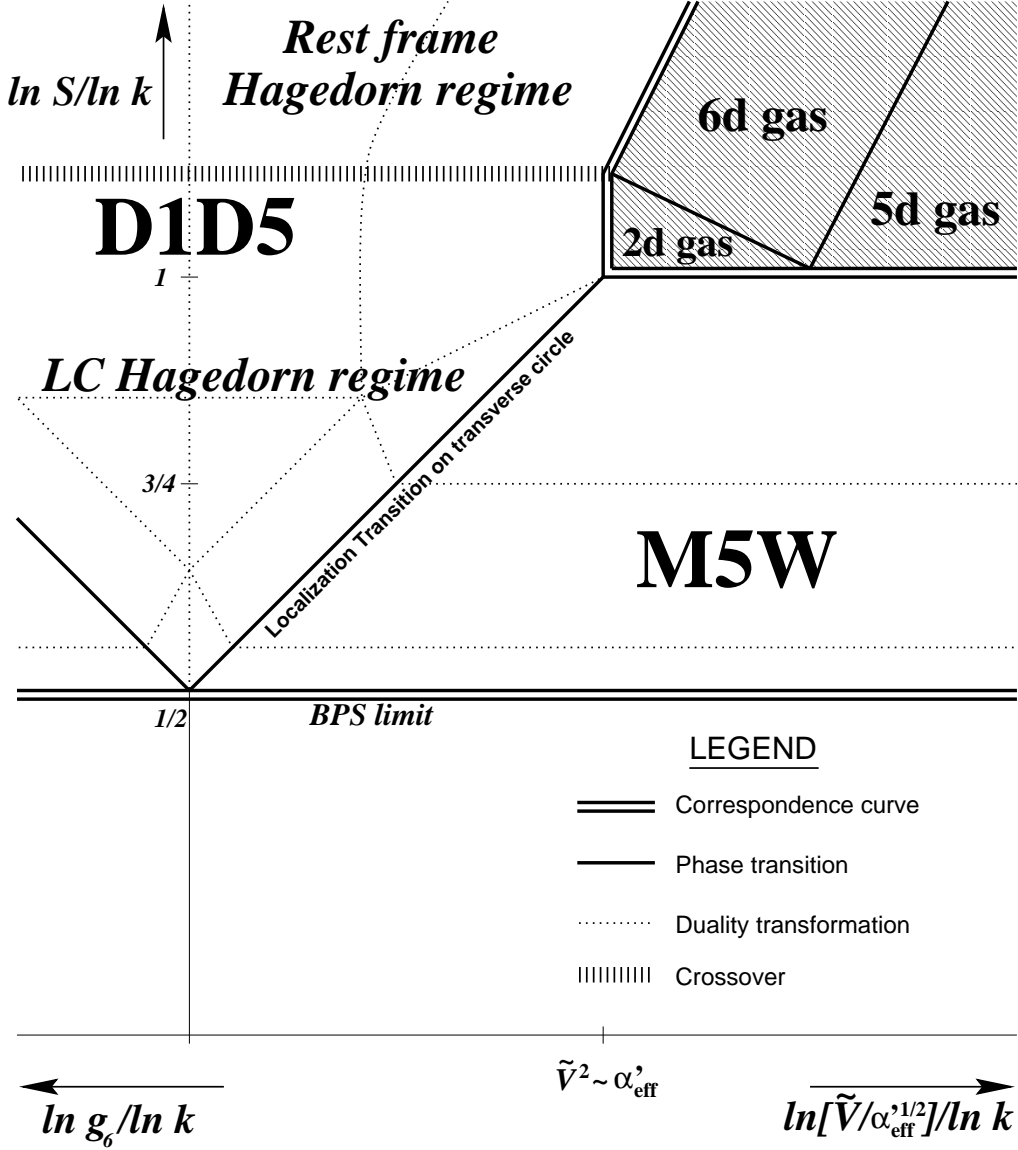}}
\caption{\sl The thermodynamic phase diagram of the little string theory
for $\Rtil\ll \sqaleff$ 
and $q=Q_1/Q_5\gg 1$; we have defined
$k\equiv Q_1 Q_5$, and consider the energy regime
of Equation~\pref{seiblim}.
The horizontal axis can be thought of either as the six-dimensional
string coupling $g_6$, or as the effective transverse box size
$\Vtil/\sqaleff=(g_6\sqrt{k})^{-1/2}$.
The thermodynamic phases described by black geometries
are labelled as follows: D1-D5 for the black D1-D5 system; 
and $M5W$ for boosted five branes whose horizon is
localized on a transverse circle.   The Hagedorn transition
of the little string theory is the horizon delocalization transition
from the viewpoint of black geometry.
}
\label{D1D51fig}
\end{figure}

\paragraph{Phase diagram for $\Rtil\ll \sqaleff$:}
Figure~\ref{D1D51fig} shows the thermodynamic phase diagram of the theory; 
the reader is referred to~\cite{MSSYM123,MSFIVE} for details
on how to construct such a diagram. 
Consider the full form of the
equation of state for the D1-D5 black geometry. The
energy above extremality is given by
\bb\label{energy}
E=\frac{R}{2g_6^2} \frac{r_0^2}{\alp^2} \lk( 1+ h_1 + h_5\re)-M_{BPS}\ ,
\ee
where $r_0$ is the location of the horizon, $M_{BPS}$ is the BPS mass
of the system
\bb
M_{BPS}=\frac{R}{2g_6^2\alp^2} \lk(\rho_1^2+\rho_5^2\re)\ ,
\ee
and
\bbb
h_{1,5}&\equiv& \lk( 1+\frac{\rho_{1,5}^4}{r_0^4}\re)^{1/2}\ ,\nonumber\\
\rho_1^2&\equiv& \frac{2g_6 \alp}{v^{1/2}} Q_1\ ,\label{h1h5}\\
\rho_5^2&\equiv& 2 g_6 v^{1/2} \alp Q_5\ .\nonumber
\eee
The entropy is given by
\bb\label{entropy}
S=\frac{2\pi R}{2g_6^2} \frac{r_0^2}{\alp^2} r_0 \lk(1+h_1\re)^{1/2}
\lk(1+h_5\re)^{1/2}\ .
\ee
Given that the metric in the string frame has 
the standard relation between the time and radial components
$g_{tt}=1/g_{rr}$, the
UV-IR relation is $E\propto r_0/\alp$.  We then consider in~\pref{energy} 
and~\pref{entropy} the energy regime~\pref{seiblim}
while holding $r_0/\alp$ fixed;
this changes only the function $h_5$ associated to the fivebranes --
the hierarchy of scales is $\rho_5\gg r_0,\rho_1$, and in the
scaling limit one drops the constant term in $h_5$.
Equations~\pref{energy} and~\pref{entropy} determine
the energy above extremality as a function of the entropy.
It takes the form of a relativistic dispersion relation
\bb\label{maineos}
E=-p_{||}+\sqrt{(S/2\pi{\sqaleff})^2+p_{||}^2}\ ,
\ee
with the longitudinal momentum
\bb\label{longmomentum}
p_{||}= \frac{k}{\tilde R}\ ;
\ee
in other words, the system can be interpreted as having $k$ units of 
longitudinal momentum on a (dual) circle of size $\tilde R=\aleff/R$.
As evident from equation~\pref{maineos},
the system behaves as a canonically 
boosted Hagedorn gas, with the invariant mass
of thermal excitations given by
\bb\label{mass}
M=\frac{S}{2\pi \sqaleff} \ ,
\ee
which is that of a string with tension $T_{ls}$.
The Hagedorn temperature is given by the effective string scale
$T_{\rm Hag}\sim 1/\sqaleff$.
From the geometrical side, the issue 
of whether the boost or the rest mass dominates
correlates directly to the horizon
radius being much smaller or much larger than the
one-brane charge radius $\rho_1$.

The bulk phase labelled D1-D5 on Figure~\ref{D1D51fig} and bounded
by the solid lines is described by equation~\pref{maineos}. Within this
phase, we have two asymptotic regimes corresponding to 
a boosted Hagedorn phase at low entropies 
and a rest frame Hagedorn phase at high entropies. 
The wide dashed line running through this phase 
denotes the crossover at 
\bb\label{hagcurve}
  S\sim \sqaleff p_{||}\sim k\; \frac{\sqaleff}{\Rtil}
\ee
between these two regimes. 
%The scaling of this curve is given by
%\bb\label{hagcurve}
%S\sim k^{3/4} g_6^{-1/2} \lk(\frac{q^{1/4}R}{V_4^{1/4}}\re)=
%	k\frac{\sqaleff}{\Rtil}\ .
%\ee
The standard decoupling limit~\pref{maldalimit} corresponds
to sending this curve to infinity, scaling out the rest frame Hagedorn
region. The transition curve~\pref{hagcurve} meets the 
vertical correspondence curve at
$g_6\sqrt{k}=\aleff/\Vtil^2\sim 1$ 
provided $\Rtil\ll \sqaleff$,
as can be seen from Figure~\ref{D1D51fig}. 
This phase diagram is then valid for $\Rtil\ll \sqaleff$.  
As we will show in the next section,
for $\Rtil\gg \sqaleff$, the scaling of the localization curve 
starts to change, and new phase structure emerges. 

The upper right corner of the phase diagram is dominated by weakly
coupled gases. For increasing $S>k$ and sufficiently large $V_4$, 
we first encounter a weakly coupled five-dimensional gas of $Q_5^2$
degrees of freedom,%
\footnote{Recall that a $d+1$ dimensional 
weakly coupled gas has the equation of state 
\bbb
E\sim S^{d+1/d}\left(c\;V_d\right)^{-1/d}\ ,
\nonumber
\eee
where $c$ is the number of degrees of freedom and $V_d$ is the
spatial volume.}
with equation of state scaling as
\bb\label{fivedgaseos}
  E\sim \frac{S^{5/4}}{(Q_5^2V_4)^{1/4}}\ .
\ee
Further increasing $S$, the temperature eventually reaches $T\sim 1/R$,
at which point the gas dynamics becomes six-dimensional
\bb\label{sixdgaseos}
E\sim \frac{S^{6/5}}{\lk(Q_5^2 R V_4\re)^{1/5}}\ .
\ee
The lower left corner of the weakly-coupled gas regime
consists of a two dimensional gas of $k$ degrees of freedom
on a circle of size $R$ with the energy scaling as $E\sim S^2/k R$
(\ie\ as in~\pref{maineos} with the boost dominating over the entropy). 
The boundary between the six- and two-dimensional gas phases 
can be found by minimizing the energies between~\pref{maineos} and
~\pref{sixdgaseos} in the boost-dominated regime of~\pref{maineos}. 
The result is
\bb\label{6d2d}
  S\sim k\; \frac{\aleff}{\Rtil\Vtil}\ .
\ee
Finally, the correspondence curve that marks the boundary
between the Hagedorn regime and the six dimensional gas 
is found to scale as
\bb\label{Hagsixcorr}
S\sim k\;\frac{\Vtil^4}{\Rtil {\alpha^{\prime\,3/2}_\eff}}
\ .
\ee
This can also be determined by minimizing
the energies between~\pref{maineos} and~\pref{sixdgaseos} 
in the entropy-dominated regime of~\pref{maineos}.
%We see that
%what appears as a transition between DLCQ and rest frame Hagedorn
%scalings in the strongly coupled regime in Figure~\ref{D1D51fig}
%metamorphoses, at weak coupling on the right of the figure, to
%a transition between two dimensional and six dimensional dynamics on the
%worldvolume of the five branes.

At sufficiently low entropies ($S \lessapprox k\Vtil^2/\aleff$), 
the appropriate duality frame
for the near-horizon geometry is that of a boosted M5-brane;
there is a circle of size $\alpha'/R$ transverse to the
M5-branes, and the horizon eventually localizes along this
circle to the phase we have labelled M5W (see~\cite{MSFIVE}).
This localization transition is the geometrical manifestation
of the Hagedorn transition.
Below, we will find an attractive interpretation of the
equation of state (see equation~\pref{locM5W} and subsequent
discussion).  This strong coupling phase transition curve
meets the correspondence curve at $\Vtil\sim \sqaleff$, and $S\sim k$.
The correspondence curve then continues horizontally at $S\sim k$
toward weak coupling, marking the boundary between the 
M5W phase and weakly coupled phases.

As mentioned above,
the Maldacena limit \pref{maldalimit} arises from the little string
limit upon a further scaling that sends $\rho_1$ to infinity
relative to $r_0$, which scales away the 
rest frame Hagedorn/6d gas/5d gas region
in Figure \ref{D1D51fig} (compare to
Figure 4 of \cite{MSFIVE}).
The third (DLCQ) energy regime~\pref{matrixlimit} of longitudinal
five-branes in Matrix theory is related to the little string limit
\pref{seiblim} as follows.
The phase diagram of Figure~\ref{D1D51fig} assumes that
$q\gg 1$, \ie\ $Q_1\gg Q_5$. As we lower $q$, at $q\sim 1$, the
four-torus, as measured at the horizon in the D1-D5 phase, becomes
string scale (see~\cite{MSFIVE} for the details); 
a T-duality on this torus is required to
go beyond this point. The new thermodynamic vacuum is again that of
a D1-D5 system with some of the parameters modified
\bb\label{dual}
v\rightarrow \frac{1}{v}\ ,\ 
q\rightarrow \frac{1}{q}\mbox{ with }
\alp,\ g_6,\ R,\mbox{ and }k\mbox{ left unchanged.}
\ee
$Q_1$ and $Q_5$ are interchanged, while the four-torus 
in string units gets inverted. We then have
\bb
v^{1/2} \alp \rightarrow \frac{\alp}{v^{1/2}}\ ,
\ee
\ie\ the limit~\pref{seiblim} where one holds the little string
tension fixed, is exchanged with the limit~\pref{matrixlimit} 
of Matrix theory; $Q_5$ is now interpreted as the boost.
In the limit, one still finds the little string theory;
one is merely reinterpreting its parameters (as is usual
in Matrix theory).
In the new variables, the fivebrane 
function $h_5$ of~\pref{h1h5} is not simplified
as before, rather the hierarchy of distance scales is
$\rho_1\gg \rho_5,r_0$, and it is the onebrane function $h_1$ that
loses its constant term in the scaling limit.
The interchange of one-brane and five-brane charges leads
(after a further S-duality) to a decoupled little
string theory whose tension is set by the NS5-brane 
before the map \pref{dual}.

\begin{figure}
\epsfxsize=16cm \centerline{\leavevmode \epsfbox{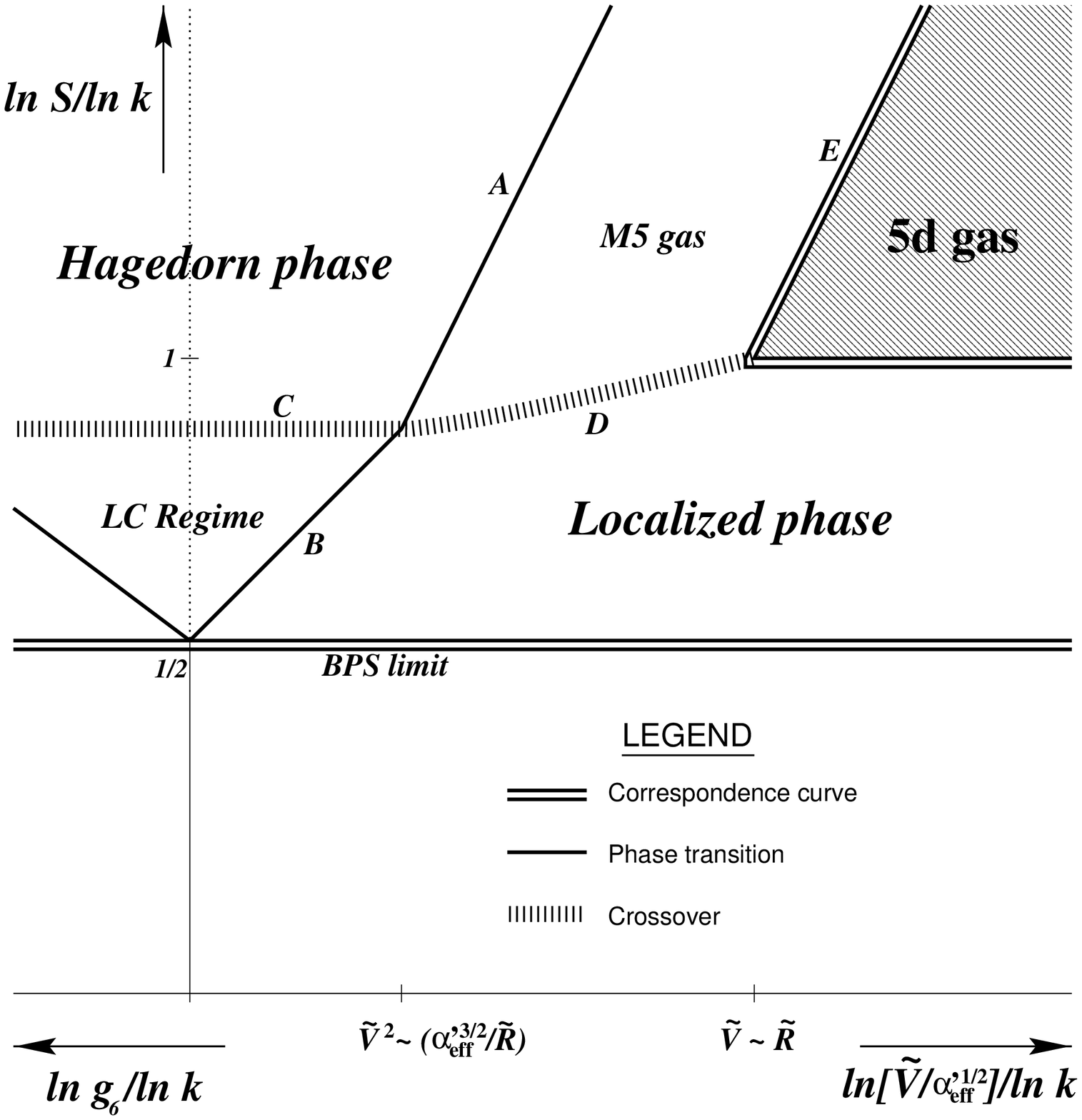}}
\caption{\sl The thermodynamic phase diagram of the 
little string theory for $\Rtil\gg \sqaleff$ and $q\gg 1$.
}
\label{D1D52fig}
\end{figure}

\paragraph{Phase diagram for $\Rtil\gg \sqaleff$:}
In this section, we chart the phase diagram for the regime 
$\Rtil\gg \sqaleff$, where
the crossover infringes upon the localization transition.
The diagram is shown in Figure~\pref{D1D52fig}.
The structure of the smeared phase describing a boosted
Hagedorn gas is exactly as before; new structure
appears in the localized phase.
The energy above extremality in the localized M5W
region in the little string limit~\pref{seiblim} is given by
\bb\label{localizedenergy}
e=y^3\lk(\frac{2}{3}-\frac{z^3}{y^3}+\lk(1+\frac{z^6}{y^6}\re)^{1/2}\re)\ ,
\ee
with the definitions
\bbb
e&\equiv& \frac{(2\pi)^6 g_6^2}{R^2} E\ ,
\nonumber\\
z^3&\equiv& (2\pi)^6 Q_1 \frac{g_6}{R v^{1/2} \alp}\ ,
\\
y&\equiv& \frac{r_0}{\alp}\ .
\nonumber
\eee
The entropy is given by
\bb\label{localizedeos}
s=y^{5/2} \lk(1+\lk(1+\frac{z^6}{y^6}\re)^{1/2}\re)^{1/2}\ ,
\ee
in terms of the quantity 
\bb
s\equiv \frac{3}{2} (2\pi)^{9/2} \frac{g_6^2}{R^{3/2}\sqaleff } S\ .
\ee
The equation of state 
above the crossover at $s\sim 1$ takes the form
\bb\label{neweos}
E\sim S^{6/5}\lk( \frac{R}{g_6 Q_5^3 V_4^{3/2}}\re)^{1/5} 
	=S^{6/5}\left(\frac{1}{k\Rtil\Vtil^4}\right)^{1/5}
\ .
\ee
We will call this phase the {\it M5 gas}.
In the first form of the equation of state, the
scaling appears to be that of a
six dimensional gas with $O(Q_5^3)$ degrees of freedom.
However, the second way of parametrizing this energy
suggests a somewhat more conventional number of degrees of freedom $k$,
in a theory whose effective string tension/coupling~\pref{efftension} is
rescaled by $Q_5$; the longitudinal circle has effective size 
$\tilde R$ ({\it c.f.} equation~\pref{dualr}),
and the transverse torus has effective size $\Vtil$
({\it c.f.} equation~\pref{veff}).
The equation of state~\pref{localizedenergy}, \pref{localizedeos}
below the crossover at $s\sim 1$ reduces to 
\bb\label{locM5W}
  E\sim \frac{g_6}{k^{3/2} R} S^3
	=\frac{1}{p_{||}}\left[
	\frac{S^{3/2}}{(k\Vtil^2)^{\frac12}}\right]^2
\ ,
\ee
having the appearance of a weakly coupled 2+1d gas
which has been boosted such that infinite momentum frame kinematics
applies.  This gas also has of order $k$ degrees of freedom
and lives in a box of linear dimension $\Vtil$.
Note that, unlike the smeared phase of equation~\pref{maineos},
the M5W phase localized on the transverse cycle
{\it does not boost canonically}.

The correspondence point for the M5 gas phase, \ie\ the point where the
curvature scale at the horizon becomes string scale, is given by
\bb\label{gfour}
S\sim k\;\left(\frac{\Vtil}{\Rtil}\right)^4
\quad,\qquad T\sim\frac1{\tilde R}\ ;
\ee
one again meets the scale of the dual circle $\tilde R=\aleff/R$.
This curve meets the crossover transition curve at $S\sim k$; 
beyond that, the correspondence point becomes the line $S\sim k$ as before. 
The phase on the other side of this correspondence boundary 
is that of the five dimensional weakly coupled gas
of $Q_5^2$ degrees of freedom, 
with equation of state scaling as~\pref{fivedgaseos}.
Minimizing the energy~\pref{neweos} 
with respect to~\pref{fivedgaseos} 
yields the transition curve~\pref{gfour}.
At temperatures less than
$T\sim (q/V_4)^{1/4}=\Vtil^{-1}$, the five dimensional 
gas freezes its dynamics on the four-torus at $S\sim k$ as in
the previous case $\Rtil\ll \sqaleff$.
The effective coupling is dressed by the size of the torus and becomes
of order one at $S\sim k$, while
the effective volume of the four-torus appears again
dressed by $q=Q_1/Q_5$.
In some cases, such effective box sizes can be interpreted as 
resulting from the typical holonomies generated by
the dynamics~\cite{MALDASUSS,BFKS1}.
It would be interesting to find a physical mechanism for
the appearance of the factor of $q^{1/4}$ in $\Vtil$.

The crossover transition from~\pref{neweos} to ~\pref{locM5W} 
appears as an extension of the transition in which
the weakly coupled gas freezes into its zero-modes; moreover,
in the low-entropy M5W phase, there is no transition as we
move toward strong coupling while staying below the crossover.
This suggests that the strong-coupling crossover in the
localized phase signals an analogous transition, in which
the M5 gas freezes into zero-mode excitations of the fivebranes.
Assuming infinite momentum frame kinematics, the invariant mass 
of these excitations is read off equation~\pref{locM5W} and
yields the equation of state of a 2+1d gas:
\bb
  M\sim S^{3/2}\left(\frac1{k\Vtil^2}\right)^{1/2}\ .
\ee
It may be that the system can find the largest amount
of available phase space by first creating a membrane
embedded in the fivebrane along the transverse cycles;
and then populating that membrane with a gas of
quasi-particle excitations.
It is interesting that the LC matrix string which 
dominates the effective dynamics in the smeared phase,
appears to become a {\em LC matrix membrane} in the localized phase.
Of course, the LC string of the former phase {\it is}
a membrane wrapped over the circle of the compactification
transverse to the M5-brane (in the M-theory duality frame
appropriate to this regime);
one might imagine that the energetics
requires this membrane to transfer its winding to cycles
along the M5-brane as the entropy is lowered.
What is seen as a transition between LC and rest frame kinematics
of a Hagedorn string in the smeared phase, is seen 
in the localized phase 
as a transition between a six dimensional gas 
of $k$ degrees of freedom in a box of size $\Rtil\Vtil^4$
and a boosted three dimensional gas of $k$ degrees of
freedom in a box of size $\Vtil^2$.

Motivated by the form of the equation of state having the appearance
of a system with $k$ degrees of freedom,
we can summarize the scaling of the various transition curves
labelled on Figure~\ref{D1D52fig} in terms of the entropy
per degree of freedom $S/k$ in the system. 
Minimizing the free energy among the various equations of state
we encounter on this diagram verifies the various phase
transition curves determined from geometrical considerations
(such as the Gregory-Laflamme localization transition and 
the correspondence principle):
\bbb
%\mbox{Curve A:}& &  
%	\frac Sk~\sim~ 
%		\left(\frac{\Vtil^4}{\Rtil{\alpha^{\prime\,3/2}_\eff}}\right)
%\ .  \nonumber \\
\mbox{Curve A:}& &  
	\frac Sk~\sim~ 
		\left(\frac{\Rtil\Vtil^4}{\alpha^{\prime\,5/2}_\eff}\right)
\ .  \nonumber \\
\mbox{Curve B:}& &  
	\frac Sk~\sim~ \left(\frac{\Vtil^2}{\aleff}\right)
\ .\nonumber \\
\mbox{Curve C:}& &  
	\frac Sk~\sim~ \left(\frac{\sqaleff}{\Rtil}\right)
\ .\nonumber \\
\mbox{Curve D:}& &  
	\frac Sk~\sim~ 
		\left(\frac{\Vtil}{\Rtil}\right)^{2/3}
\ .\nonumber \\
\mbox{Curve E:}& &  
	\frac Sk~\sim~ 
		\left(\frac{\Vtil}{\Rtil}\right)^{4}
\ .\nonumber
\eee
The choice of parametrization of the thermodynamics
at strong coupling in terms of the five independent variables 
$S$, $\tilde R$, $\Vtil$, $\aleff$, and $k$
is more than a mere reshuffling of the original parameters
$S$, $R$, $V_4$, $g_6$, $k$, and $q$, which are six
in number (the seventh parameter, $\alp$, is scaled out in the decoupling
limit); the ratio $q=Q_1/Q_5$ has disappeared from
the strong coupling equations of state
~\pref{maineos}, \pref{neweos}, and \pref{locM5W}.
This reduction of parameters indicates that we have found the
proper interpretation of the thermodynamic data.
Furthermore, the uniform linear scaling of the transition
curves in $S$ with respect to $k$ (see also~\pref{Hagsixcorr}) supports
the interpretation of $k$ as the number of degrees of freedom in the system.
The fact that these curves are parametrized by dimensionless 
cycle sizes $\Rtil/\sqaleff$ and $\Vtil/\sqaleff$,
with no other dependence on the combinations of charges $q$ or $k$,
indicates a nontrivial scaling in the thermodynamics.

We hope this interpretation of the thermodynamics
leads to further progress in understanding the dynamics of fivebranes.
There seems to be a remarkable parallel with the Matrix model
of M-theory.  It is hard to understand what are the appropriate
degrees of freedom to use in formulating M-theory in flat space;
adding charge to the system in the form of $N$ units of momentum
along a circle allows a truncation of the dynamics of the theory to 
$N^2$ degrees of freedom (the D0 brane sector).
In the case of a system of $Q_5$ fivebranes, it is hard to understand
how the apparently $O(Q_5^3)$ degrees of freedom originate.
Again adding another charge to the system in the form
of $Q_1$ units of momentum,%
\footnote{In the duality frame appropriate to the M5 gas,
$Q_1$ is a momentum rather than a winding quantum 
as in the original D1-D5 duality frame.}
one achieves an effective
description of the system in terms of $O(k=Q_1Q_5)$
degrees of freedom.

\paragraph{\bf Acknowledgments:}
This work was supported by DOE grant DE-FG02-90ER-40560.

\providecommand{\href}[2]{#2}\begingroup\raggedright\endgroup

\end{document}